\documentstyle[sprocl,epsf]{article}

\def\ep{\epsilon}
\def\si{\sigma}

\def\ra{\rightarrow}

\def\al{\alpha}

\def\be{\begin{equation}}
\def\ee{\end{equation}}
\def\bea{\begin{eqnarray}}
\def\eea{\end{eqnarray}}
\newcommand{\defect}{{\rm\small defect}}
\newcommand{\matter}{{\rm\small matter}}
\newcommand{\eq}{{\rm eq}}
\newcommand{\de}{\delta}
\newcommand{\De}{\Delta}
\newcommand{\la}{\lambda}
\newcommand{\ga}{\gamma}
\newcommand{\Ga}{\Gamma}
\newcommand{\dd}{\partial}
\newcommand{\MM}{{\cal M}}
\newcommand{\gsim}{\stackrel{>}{\sim}}
\newcommand{\lsim}{\stackrel{<}{\sim}}

\begin{document}
\title{Signatures of Topological Defects in the Microwave Sky: An 
	Introduction}

\author{ R. Durrer}

\address{Univ\'ersit\'e de Gen\`eve, D\'epartement de Physique Th\'eorique\\
24, Quai E. Ansermet\\ CH--1211 Gen\`eve, Suisse}


\maketitle
\abstracts{An introduction to topological defects in cosmology is
given. We discuss their possible relevance for structure
formation. Especial emphasis is given on the signature of topological
defects in the spectrum of anisotropies in the cosmic microwave
background. We present simple analytic estimates for the CMB spectrum
on large and intermediate scales and compare them with the
corresponding approximations for models where initial perturbations
are generated during an inflationary epoch.}
  
\section{Introduction}

The formation of structure in the universe is one of the mayor open
problems in cosmology. Already in 1946 Lifshitz has noted \cite{lifshitz}
that expansion counteracts gravitational attraction in such a way,
that in an expanding universe the gravitational potential cannot grow
by linear gravitational instability. Also the growth of density
perturbations is reduced to a power law due to expansion. In a
radiation dominated universe, radiation pressure inhibits any significant
growth of density fluctuations. In all, density fluctuations can have
grown at most by a factor $\sim a_0/a_\eq = z_\eq \sim 10^4$ due to linear
gravitational instability, where $a_0$ denotes the value of the
cosmological scale factor today and  $a_\eq$ denotes its value at the
time of equal matter and radiation.

Therefore, initial fluctuations on the order of $10^{-4}$ caused by some 
other mechanism than gravitational instability are needed. Such
initial fluctuations can then be enhanced by gravity
 and lead to density fluctuations of order unity and
finally to the observed structures in the universe.
Currently two classes of models which can generate initial perturbations
are under investigation. 

In the first class, initial perturbations
emerge from quantum fluctuations of a scalar field, which expand during a
period of inflation to scales larger than the Hubble scale and then ``freeze
in'' as classical fluctuations in the energy density. Generically,
inflationary models lead to a so called Harrison-Zel'dovich spectrum
of fluctuations\cite{haze}. This spectrum is defined by the
requirement of having constant mass fluctuations at horizon crossing:
\be
\left \langle\left({\de M\over M}\right)^2_{k_H(t),t}\right\rangle 
	= \mbox{ constant,} 
\ee
where $k_H(t)=2\pi/t$ denotes the wave number corresponding to the
horizon scale at time $t$.

In inflationary models, initial fluctuations typically are Gaussian,
{\em i.e.}, with random initial phases. After inflation they evolve
deterministically according to  homogeneous linear cosmological
perturbation equations. The evolution of an arbitrary mode $k$ of a
perturbations variable $\delta$ can thus be described by means of a 
deterministic transfer function $T$ and the initial value 
$\delta(t_i)$ is perfectly coherent with
$\delta(t)=T(t,t_{i})\de(t_{i})$. In other words
\be
{\langle|\delta(t_i)\de(t)|\rangle\over
\sqrt{\langle|\de(t_i)|^2\rangle\langle|\de(t)|^2\rangle}}=
{\langle|\delta(t_i)T(t,t_i)\de(t_i)|\rangle\over
\sqrt{\langle|\de(t_i)|^2\rangle\langle|T(t,t_i)\de(t_i)|^2\rangle}}=1~.
\label{cohe}
\ee

In the second class of models, perturbations are induced by 
topological defects which may form during a symmetry breaking phase
transition. This mechanism is explained in the next section. The
amplitude of initial fluctuations due to topological defects which
form at a symmetry breaking energy scale $\eta$ is on the order of
$\ep=4\pi G\eta^2$.  To obtain the correct amplitude thus requires
defects which form during a  phase transition at GUT scale 
$\eta\sim 10^{16}$GeV.

In this situation, perturbations in the cosmic fluid are constantly 
{\em sourced}
by topological defects and evolve according to {\em inhomogeneous}
linear perturbation equations. Since the defects make up only a small
perturbation of the cosmic energy density and since (soon after the phase
transition) they do not interact with the cosmic fluid other than
gravitationally, they evolve according to the unperturbed geometry (in
linear perturbation approximation). However, defect evolution is in
general non-linear and the random initial conditions of the source
term in the linear cosmological perturbation equations of given scale
$k$ 'sweep' into other scales. Therefore, the perfect
coherence of inflationary perturbations is no longer maintained
and Eq.~(\ref{cohe}) is violated. How strong and how significant
this decoherence is, depends on the details of the model considered.

Due to this general behavior inflationary models are sometimes called
'coherent' and 'passive' (no source terms in the linear perturbation
equation) while defect models are called 'decoherent' and 'active'
(fluid perturbations are constantly sourced by the defect energy
momentum tensor) \cite{joao}. In this talk, we concentrate on the 
second class of models. The main similarities and differences of the 
two classes are summed up in table~1. 
\begin{table}[htb]
\begin{tabular}{||l| l||}
\hline\hline
Inflationary models & Topological defects\\
\hline
\multicolumn{2}{||c||}{Similarities} \\ \hline
\multicolumn{2}{||l||}{$\bullet$ Cosmic structure formation is due to
	gravitational instability of small} \\ 
\multicolumn{2}{||l||}{ ~'initial' fluctuations. $\rightarrow$ 
	Gravitational perturbation theory} \\
\multicolumn{2}{||l||}{~can be applied.}\\
\multicolumn{2}{||l||}{$\bullet$ GUT scale physics is involved in
	generating initial fluctuations.}\\
\multicolumn{2}{||l||}{$\bullet$ The only relevant 'large scale' is
	the horizon scale. $\rightarrow$ Harrison-} \\
\multicolumn{2}{||l||}{~Zel'dovich spectrum.}\\
\hline
\multicolumn{2}{||c||}{Differences} \\ \hline
$\bullet$The amplitude of fluctuations &$\bullet$The amplitude of 
fluctuations\\
~depends on details of the inflatio-&~is fixed by the symmetry breaking \\
~nary potential $\rightarrow$ fine tuning.&~scale $\eta$,
$\epsilon=4\pi G\eta^2$.\\
$\bullet$The linear perturbation eqs. are&$\bullet$The linear perturbation 
eqs. are in-\\
~homogeneous (passive). &~homogeneous, have sources
(active).\\
$\bullet$For given initial perturbations, &$\bullet$The source evolution is 
 non-linear\\
~the entire problem is linear. &~at all times.\\ 
$\bullet$Randomness enters only in the  &$\bullet$Randomness enters
at all times due\\~initial conditions. &~to the mixing of scales in the
non-\\ &~linear source
evolution (sweeping).\\
$\bullet$The phases of perturbations at &$\bullet$Phases may become incoherent.\\
a given scale $\lambda$ are coherent. &\\
$\bullet$There exist correlations on super&$\bullet$No correlations on
super Hubble \\
~Hubble scales.&~scales.\\
~Perturbations are 'acausal'. &~Perturbations are 'causal'.\\
\hline\hline
\end{tabular}
\caption{ Similarities and and differences of inflationary
perturbations versus perturbations induced by seeds.}
\end{table}
\newpage

\section{Topological Defects}
Topological defects are as ubiquitous in physics as are symmetry
breaking phase transitions. 
Usually they are described by means of a scalar field (order
parameter, Higgs field) evolving in a temperature dependent
potential. In the Landau Ginzburg theory of super-conductivity, {\em e.g.},
the order parameter represents the ``Cooper pairs'' which are
described by means of a complex scalar field. In this example, the
scalar field is electrically charged and interacts with the
electromagnetic gauge field.  For sake of simplicity, we consider here
a pure scalar field $\phi$, with $\phi^4$ interaction term but
 without gauge field.  If $\phi$
is in a thermal bath at temperature $T$ and we have 'integrated out' the
excitations of energies $E\ll T$, we obtain an effective Lagrangian
density with  temperature dependent potential \cite{Kaputza,Kibble}
\be
{\cal L}(\phi) = {1\over 2}(\dd_\mu\phi)^2 -V_T(\phi) ~.
\ee
At very low temperature $V$ approaches the zero temperature potential,
$V_0={1\over 4}\la(|\phi|^2-\eta^2)^2$ with vacuum manifold
\be
{\cal M}_0=\{\phi\left| |\phi|^2=\eta^2\} \right. ~.
\ee
(The vacuum manifold denotes the space of minima of the potential
$V$.)
At higher temperatures, there are corrections to $V$ which in general
depend on the interactions of the scalar field with other (fermionic
and bosonic) fields. In our simple case, the main correction is of the
form $T^2\phi^2$ which, at high enough temperature, namely for
$T>T_c=2\eta$, changes the 'Mexican hat' into a parabolic shape with
$\MM_0=\{0\}$. 
\begin{figure}[htb]
\vspace{0.5cm}
\centerline{
\epsfysize=5.5cm
\epsffile{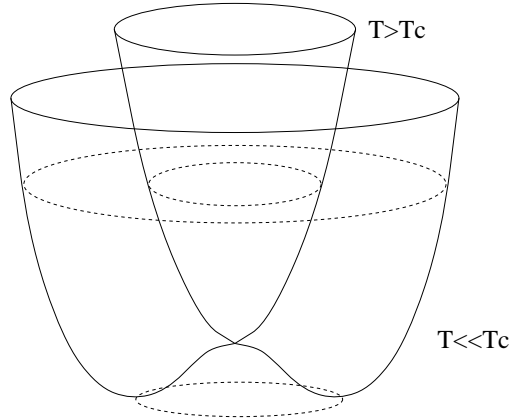}}
\caption{The temperature dependence of potential of a complex scalar
field. The vacuum manifold is a circle for $T<T_c$ and a point for
$T>T_c$.}
\end{figure}
At $T>T_c$ therefore, not only the Lagrangian density but also the
only possible vacuum state $<\phi>=0$ is symmetric under phase
rotations, $\phi\ra e^{i\alpha}\phi$. As soon as the temperature falls
below $T_c$, the vacuum manifold becomes a full circle, $\MM_0 ={\bf
S}^1$ and a given vacuum state  $<\phi>=r(T)e^{i\alpha}$ is no longer
invariant under phase rotations. The function $r(T)$ is a temperature
dependent amplitude with 
\be
r(T_c)=0 \mbox{ and }~~~~ \lim_{T\ra 0} r(T)=\eta~.
\ee
This process is the 'spontaneous' breakdown of a symmetry (here phase
rotations or $U(1)$). Even though the Lagrangian density and $\MM_0$  
as a whole are invariant
under phase rotation, at $T<T_c$, this is no longer manifest in phenomena
which can be described by expansion around the vacuum, since the
choice of a vacuum state spontaneously breaks the symmetry.

If such a phase transition takes place in the early universe, the
coherence length is finite (it is bounded by the causal
horizon). As the universe cools below the critical temperature,
 we expect the field $\phi$ to assume different
vacuum expectation values at different patches of space which are separated by
distances larger than the coherence length.  If we now prescribe a
closed curve in a plane through physical space,
 $\ga: [0,1]\ra {\bf R}^2: s\ra {\bf x}(s)$, ${\bf x}(0)={\bf x}(1)$,
the field $\phi$ can change its phase along the curve
$\Ga=\ga([0,1])$. $<\phi>({\bf
x}(s))=r(T)e^{i\alpha(s)}$. By continuity reasons the phase must
change by a multiple of $2\pi$ during a full turn,
$\alpha(1)=\al(0)+n2\pi$. If $n\neq 0$, the loop $\Ga$ cannot be
contracted to a point with $\al(s)$ changing continuously. The 
expectation value $<\phi>$ thus must pass through 0 somewhere in 
the interior of $\Ga$. In other words, $\phi$
must leave the vacuum manifold and assume a state of higher energy in
some small region in the interior of $\Ga$.
\begin{figure}[htb]
\centerline{
\epsfysize=4.5cm
\epsffile{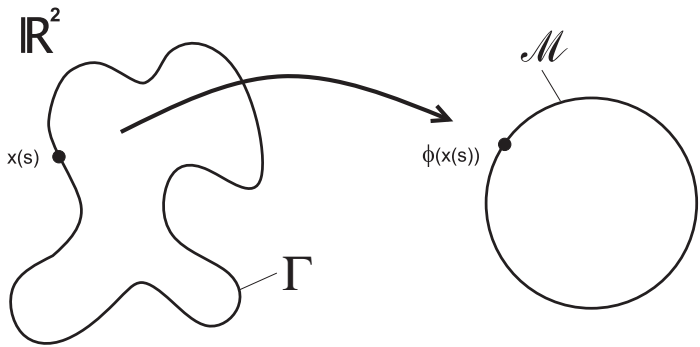}}
\caption{If the map indicated in this figure,
	$:\Ga\ra \MM: {\bf x}(s)\mapsto \phi({\bf x}(s))$ is 
	non-trivial, a cosmic string forms.}
\end{figure}
 If we now leave the plane
and continue this argument in the third dimension, we obtain a long,
thin string within which $\phi$ assumes higher energy, a cosmic
string. The cross section of the string, {\em i.e.} of the region where 
$\phi$ leaves the vacuum manifold, is typically of order $1/\eta$. The
length of a string is either infinite or the string must be closed.
The mechanism of defect formation described here is called the
 'Kibble mechanism' \cite{Kibble}.

The main ingredient for the Kibble mechanism is  the existence non-trivial
(non-shrinkable) maps from a closed curve is space $\sim {\bf S}^1$ to
the vacuum mani\-fold. The classes of these maps form a group, the
first homotopy group $\Pi_1(\MM)$.

Similarly, if maps from spheres in space into $\MM$ cannot be shrunk
to a point, {\it i.e.} $\Pi_2(\MM)$ is non-trivial, $\phi$ might have
to leave the vacuum mani\-fold in a small patch in 3-space, leading to a
tiny region of higher energy, a monopole. Again, the spatial extension
of the monopole is on the order of $1/\eta$ and thus extremely small 
in comparison to cosmological scales.

Furthermore, if we consider configurations which are asymptotically
constant ($\phi({\bf x})\ra_{|{\bf x}|\ra\infty}$ const.), 
we can compactify 3-space to $\overline{{\bf R}^3}=({\bf
R}^3\cup\infty) \sim {\bf S}^3$, and assign to $\phi(\infty)$ the value
of the asymptotic constant. We then encounter the question whether
there exist non-shrinkable maps from ${\bf S}^3\ra \MM$. (Non-trivial
homotopy group $\Pi_3(\MM)$.) One can show,
that such a configuration is always unstable and will shrink and
eventually leave the vacuum mani\-fold and unwind. (In the case of
finite energy configurations this is Derrick's theorem\cite{Derrick}.) A
configuration $\phi$ which winds once around ${\bf S}^3$ 
 is called 'texture' of winding number 1. (Textures of higher winding
numbers are probably unstable and decay into simple textures.)

There are some doubts about the applicability of this concept
to cosmology; especially the notion of an asymptotically constant
configuration is not at all causal. However, in the case of $\Pi_3$
(and only in this case!) one can define a texture number density
$n_\phi(x)$, such that
\be w_\phi({\bf R}^3)=\int_{{\bf R}^3}n_\phi(x)d^3x  \ee
determines the winding number ({\em i.e.} texture number) of the map
$\phi:\overline{{\bf R}^3}\ra\MM$. Clearly, if this map is well
defined, $w_\phi({\bf R}^3)$ is always an integer. Nevertheless, 
we can also
consider the winding number in a finite volume $V\in {\bf R}^3$ and 
determine  
\be w_\phi(V)=\int_Vn_\phi(x)d^3x  \ee
which need not be an integer. Numerical investigations\cite{Julian} 
have shown that
a configuration shrinks whenever $w_\phi(V_H)\gsim 1/2$, where $V_H$
denotes the Hubble volume,  independent of the behavior of $\phi$ at 
spatial infinity. 
Therefore, it makes sense to talk about textures also in a
cosmological context.

Depending on whether the symmetry is local (gauged) or global
(rigid),  defects are called 'local' or 'global'. In the
case of local defects, gradients are compensated by the gauge
potential ($\dd_\mu \ra D_\mu=\dd_\mu+ieA_\mu$), and there is no
considerable gradient energy. This has two important consequences:
\begin{itemize}
\item The energy of defects is
strongly confined, {\em i.e.} the extension of  defect energy 
is given by the inverse symmetry breaking scale, $1/\eta$.
\item Soon after their formation, local defects cease to interact.
There are no long range interactions between local defects.
\end{itemize}
In the case of global defects, there are no gauge fields to compensate
gradients and the energy is dominated by gradient energy which is
 spread out over  typically the  horizon scale $t$.
Interactions between defects are very strong. Defects of opposite charge
annihilate leading to a few (or less) defects per horizon
volume. Energy density always behaves like $\rho_{\defect}\sim
\eta^2/t^2$ (up to possible logarithmic corrections) and we thus 
find $\rho_{\defect}/\rho_{\matter} \sim 4\pi G\eta^2=\ep$. The defect 
energy amounts to a constant small fraction of the total energy 
density of the universe. This behaviour is called scaling. 

In the case of local defects, only strings scale. Local monopoles soon
come to dominate the energy density of the universe and local textures
quickly die out. Defects responsible for structure formation and CMB
anisotropies are thus either local strings or global defects.

In the case of global fields, gradient energy is the
main seed for  perturbations in the geometry. Whether these gradients
lead to topological defects or not is actually less important. 
{\em E.g.} scalar fields
with a $\phi^4$ potential and $N>4$ components do not lead to
topological defects in 3-dimensional space, but structure formation
seeded by such fields is very similar to the $N=4$ (global texture) 
and $N=3$ (global monopoles) models. The limit $N\ra \infty$ where the
field equations can be solved exactly, provides a useful
approximation to global defect scenarios \cite{TS,KD}. 

At temperatures significantly below the symmetry breaking scale, the
dimensionless field $\beta=\phi/\eta$ obeys to a very good approximation
the  scale free equation of a non-linear
$\si$--model \cite{Cam}. Scaling arguments then yield
${\cal O}(\dot{\beta})={\cal O}(\dd_i\beta)= {\cal O}(1/t)$.

The typical amplitude of geometrical fluctuations in scaling defect models is
given by 
\be 4\pi G\rho_{\defect}/(\dot{a}/a)^2 \sim 4\pi G\eta^2\equiv \ep~.\ee
The COBE experiment provides the normalization $\ep \sim 10^{-5}$ and
thus $\eta\sim 10^{16}$GeV. In order to create large enough
fluctuation to seed the formation of structure in the universe,
defects must thus form during a symmetry breaking phase transition at
$T_c\sim 10^{16}$GeV, {\em i.e.} at a typical scale of grand unification.
 
\section{The CMB Anisotropy Spectrum from Topological Defects}
Anisotropies in the cosmic microwave background (CMB) are small and can 
thus be calculated within linear cosmological perturbation theory.

If we neglect the finite thickness of the last scattering surface,
$t=t_{dec}$, the temperature anisotropies in the cosmic microwave
background can be found by
integrating photon geodesics from $t_{dec}$ until today, $t_0$.
This leads to
\be
{\delta T\over T}({\bf n}) = \left[
	 {1\over 4}D^{(r)} +V_jn^j 
	+(\Psi-\Phi)\right](t_{dec},{\bf x}_{dec})
	+ \int_i^f (\dot{\Psi} - \dot{\Phi} )({\bf x}',t') dt  ~. 
\label{dT} 
\ee
Here $D^{(r)}$ denotes a gauge invariant variable describing intrinsic
density fluctuations in the radiation, $\bf V$ is the peculiar baryon
velocity, {\em i.e.} the velocity of the emitter (the corresponding term
for the observer, which just results in the well-known dipole
contribution has been omitted), and $\Phi$ and $\Psi$ are the Bardeen
potentials\cite{Bardeen}, two geometric variables which describe scalar
perturbations of the Friedmann geometry. $\Psi$ is a close analog of
the Newtonian potential. If the matter causing the geometric
perturbation is either non-relativistic or an ideal fluid, $\Phi=-\Psi$.
A derivation of Eq.~(\ref{dT}) can be found in Ref.~12.

The first two terms in Eq.~(\ref{dT}) are mainly caused by acoustic 
oscillations of the baryon photon fluid prior to recombination. This 
causal process acts only on sub horizon scales and thus comes to dominate 
on angular scales $\theta<1^o$. The third term is the ordinary Sachs 
Wolfe contribution\cite{SW}. It is due to the gravitational potential at 
the last scattering surface, which induces a redshift(blueshift) of the 
free photons climbing out of it (falling down from it). The integral in 
Eq.~(\ref{dT}) is the integrated Sachs Wolfe term, which is induced by 
a time varying gravitational potential along the path of the photon from 
the last scattering surface into the antenna of the observer. The Sachs 
Wolfe contributions are relevant on large angular scales, $\theta\gg 1^o$.

The finite thickness of the last scattering surface leads to diffusion 
damping at very small angular scales: During recombination, the mean free 
path of photons grows from effectively $0$ to much larger than the Hubble 
scale. Perturbations with are small enough so that  photons can
diffuse out of them during the recombination process, are thus damped 
away. This process is called Silk damping\cite{Silk}. To a reasonable 
approximation it can be taken into account by multiplying the result of 
Eq.~(\ref{dT}) with an exponential damping envelope\cite{HuSu}. For a 
more accurate treatment, one has to solve Boltzmann's equation taking into 
account non-relativistic Compton scattering of photons and 
electrons\cite{Review}. 

In addition to these fluctuations which are determined entirely within
 linear perturbation 
theory, some secondary effects due to the formation of the first 
non-linear structures might influence the perturbations. There are 
notably gravitational lensing, the Rees Sciama effect and the Sunyaev 
Zel'dovich effect which can influence very small scales; as well as early
reionization which may lead to damping of fluctuations on 
intermediate scales. Here we just discuss the Sachs Wolfe and acoustic
contributions which dominate on large and intermediate angular scales.

Since ${\delta T\over T}$ is a function on the sphere, it make sense to 
expand it in terms of spherical harmonics:
\be
{\delta T\over T}({\bf n})=\sum_{\ell m}a_{\ell m}Y_{\ell m}({\bf n})
  ~. \label{harm}\ee
The anisotropy spectrum is then defined by
\be
C_{\ell}={\langle\sum_m|a_{\ell m}|^2\rangle\over 2\ell+1} ~.
\ee
In the case of Gaussian perturbations, the $C_{\ell}$'s contain the full 
statistical information of the CMB anisotropies since they are the 
'harmonic transform' of the two point correlation function\footnote{The 
expansion into spherical harmonics on the sphere is the exact analog 
of Fourier transform in ${\bf R}^n$. Since the sphere is compact, the
'harmonic transform' of a function on the sphere lives on a discrete set.}:
\be  {\cal C}(\cos\vartheta) \equiv
 \left\langle{\delta T\over T}({\bf n}){\delta T\over T}({\bf n}')
\right\rangle\left|_{{~}_{\!\!({\bf n\cdot n}'=\cos\vartheta)}}\right. =
  {1\over 4\pi}\sum_\ell(2\ell+1)C_\ell P_\ell(\cos\vartheta)~, 
\label{2point} \ee
where $P_\ell$ denotes the Legendre polynomial of order $\ell$.

Since the relevant quantity for CMB anisotropies are the $C_\ell$'s, 
angular regimes are often translated into intervals of $\ell$'s. 
Small $\ell$'s probe large angular scales whereas large $\ell$'s probe 
small angular scales. The angular scale corresponding to a given $\ell$ 
is about $\theta_\ell\sim 1/\ell$. In terms of $\ell$, 'large angular 
scales' correspond to $\ell\lsim 50$ and Silk damping becomes relevant at 
$\ell\gsim 800$. The scales in between are intermediate angular scales.
\subsection{Large scales}
 Angular scales, $\theta\gg 1^o$, which correspond to spherical
 harmonics with index $\ell\ll 200$ subtend a distance which is 
larger than the size of the horizon at recombination. Temperature
 fluctuations on these angular scales are either due to super horizon 
fluctuations, if they result from fluctuations at the last scattering
 surface, the 'recombination shell', or they have been induced during
 the propagation of the photons from the last scattering surface into the
 antenna of the observer.

The Sachs Wolfe (SW) contributions to the CMB anisotropies from inflationary 
models and defect models are as different as they can be. 
Nevertheless they finally lead to the same Harrison Zel'dovich spectrum 
of $C_\ell$'s. Let us elaborate on this 'accident' in some detail.

For a pure CDM (cold dark matter) model, it is easy to show from the 
linear perturbation 
equations that $\Phi=-\Psi$ and $\dot{\Psi}=0.$ Furthermore, assuming 
adiabatic perturbations, one finds from the 
analog of Poisson's equation,  $(1/4)D^{(r)}=-(5/3)\Psi +{\cal O}((kt)^2)$.
On super horizon scales, $kt\ll1$, Eq.~(\ref{dT}) thus yields for pure CDM
\be 
{\delta T\over T}({\bf n})={1\over 3}\Psi(t_{dec},{\bf x}_{dec})~. \label{CDM}
\ee
This is the well known Sachs Wolfe result. For a typical inflationary 
spectrum, the Bardeen potentials behave like
\be
\langle|\Psi(k)|^2\rangle\propto1/k^3 \mbox{ ~~~~ for inflation + CDM.}
 \label{k1CDM}\ee
 Using this and 
Eq.~(\ref{CDM}), one can calculate the anisotropy spectrum and finds (see 
appendix)
\be
C_\ell\propto {1\over \ell(\ell+1)}~. \label{cl}
\ee 
For topological defect models, the situation is very different. One can 
show, that (due to compensation) the Bardeen potentials have white 
noise spectra on super horizon sales \cite{joao,DS}. By dimensional 
reasons therefore $\langle|\Psi(k)|^2\rangle t^{-3}=$constant. Furthermore, 
one finds that $D^{(r)}$ behaves like $(kt)^2\Psi$ and is thus 
negligible on super horizon 
scales. Once a perturbation enters the horizon, $t\sim 1/k$ the defect 
contribution decays and it is dominated by the contribution due to CDM
which then becomes time independent. A reasonable approximation to the 
Bardeen potentials from defect models is thus
\be
\langle|\Psi(k)|^2\rangle\propto\left\{\begin{array}{lll}
 t^3, &\mbox{on super horizon scales,}& kt\le 1,\\
  1/k^3, &\mbox{on sub horizon scales,}& kt\ge 1,\end{array}
  \right.\mbox{ for defects + CDM.} \label{defects}
\ee
Using this approximation, it is easy to see that the ordinary Sachs Wolfe 
effect is very small (for scales which are super horizon at decoupling), 
whereas the integrated SW term behaves like the inflationary SW contribution 
leading to the same spectrum of CMB anisotropies on large angular scales, 
Eq~(\ref{cl}). A derivation of this result is given in the appendix.

Of course, our argumentation in the case of topological defects is very 
crude. It is, however, useful to interpret the findings from numerical 
si\-mu\-la\-tions. Large scale CMB anisotropies from simulations of 
Global defects\cite{DZ} are presented in Fig.~3. Similar results have
been obtained for cosmic strings\cite{String}.
\begin{figure}[htb]
\epsfysize=6.5cm
\centerline{
\epsffile{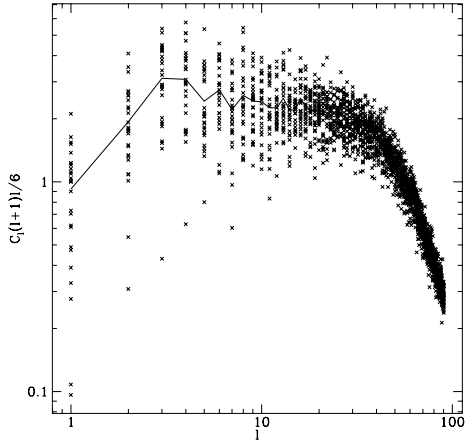}}
\caption{The large scale CMB anisotropies from global texture
simulations. The spectra for 27 observers are plotted. The sharp drop
after $\ell\sim 30$ is due to finite resolution. 
(Figure taken from Durrer and Zhou, Ref. 17.)}
\end{figure}

\subsection{Intermediate scales}
The signal from CMB anisotropies on intermediate scales is
most interesting since it contains the most structure and thus the
most detailed information.

As already mentioned, fluctuations on intermediate scales are due
to acoustic oscillations in the tightly coupled photon baryon plasma prior to
recombination. To understand the basic principle, we treat these
oscillations here in a very simple way. We neglect the presence of
baryons and thus set the sound velocity of the plasma,
$c_s^2=1/3$. Energy and momentum conservation then lead to the
following equations of motion for the density perturbation $D^{(r)}$ and
the peculiar velocity potential $V^{(r)}$ (see Ref. 12):  
\be
\dot{D}^{(r)}=- {4\over 3}kV^{(r)}~~,~~~~ \dot{V}^{(r)}=2k\Psi
	+{1\over 4}kD^{(r)} ~.
  \label{em}\ee
(In the second equation we have suppressed the difference between $\Psi$
and $-\Phi$ which is unimportant for our qualitative discussion.)
Eqs.~(\ref{em}) can be combined to a second order equation for $D$:
\be
\ddot{D}^{(r)} +{1\over 3}k^2D^{(r)}=-{8\over 3}k^2\Psi ~. \label{acou}
\ee
Using the behavior  $\Psi\propto t^{3/2}$ on super horizon scales, and
$\Psi\propto k^{-3/2}$ on sub horizon scales, we obtain the solutions
\be D^{(r)}= \left\{\begin{array}{lll}
 -{32\over105} (kt)^2\Psi &\mbox{on super horizon scales,}&kt\ll 1\\
  8\Psi(\cos(kt/\sqrt{3})-1)&\mbox{on sub horizon scales,}&
kt\gg 1.
\end{array} \right.
\ee

This behavior of $D^{(r)}$ is very different to the adiabatic 
inflationary case. 
There $D^{(r)}\sim\Psi$ on super horizon scales and $D^{(r)}\sim 
\Psi(\sin(kt/\sqrt{3})-1)$ on sub horizon scales. For defects 
thus $D^{(r)}$ is very small at horizon crossing and first has to 
grow to achieve its maximum whereas in the adiabatic inflationary 
case, $D^{(r)}$ is already at its 
maximum on super horizon scales and starts decaying at horizon crossing. 
This is the reason why the first acoustic peak is displaced to larger 
$\ell$ for defect models. In a flat, $\Omega=1$ universe the position of 
the first acoustic peak is about $\ell_{peak}\sim 360$ for global defects 
where it is at $\ell=220$ for adiabatic inflationary models\cite{CT,DGS}.

I carefully always said 'adiabatic inflationary models' since the second 
order Eq.~(\ref{acou}) of course allows for two modes and in 
inflationary models one is actually free to choose the adiabatic mode, 
which is defined by $V^{(r)}=V^{(CDM)}$ on super horizon scales or the 
isocurvature mode which is defined by $D^{(r)}\ra_{kt\ra 0} 0$. For 
defect models, however, we  want to pick out the peculiar solution 
induced by the defect fluctuations without adding an arbitrary 
homogeneous solution, a perturbation which then would have
 to be induced by some 
other mechanism like, {\em e.g.} inflation. For defect models we thus have no 
freedom in the choice of the mode. The resulting CMB anisotropy spectrum 
for a typical model with global defects is shown in Fig.~4.  

\begin{figure}[htb]
\epsfysize=6.5cm
\centerline{\epsffile{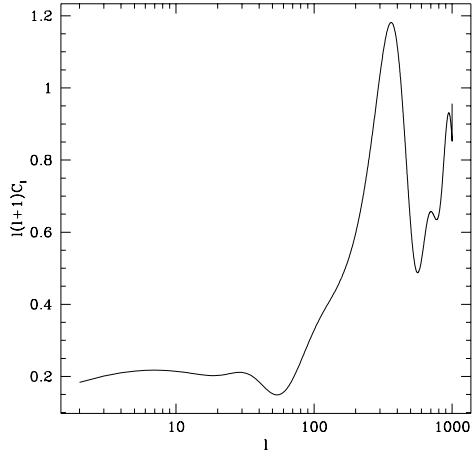}}
\caption{The CMB anisotropy spectrum from global topological
defects. The vertical scale is arbitrary. This 
result was obtained by a model calculation as explained in Ref. 16.}
\end{figure}

An additional phenomenon which can be important for the acoustic peaks in 
the CMB anisotropy spectrum is decoherence: For inflationary 
perturbations the phase of an acoustic oscillation is entirely determined 
by its wave number, {\em i.e.} all fluctuations with fixed wave number $k$ 
are at fixed phase in their temporal oscillations. In models with 
topological defects, the fluctuations 
are induced by the defect energy which evolves in a complicated non-linear 
way. Decoherence can be described by the decay of the correlation function
\be C(k,t) = {\langle|\Psi(k,t)\Psi(k,t_i)|\rangle\over\sqrt{
	\langle|\Psi(k,t)|^2\rangle\langle|\Psi(k,t_i)|^2\rangle}} ~. 
	\label{decor}
\ee
Since defects evolve causally, they are 'frozen in' on super horizon 
scales and no decoherence can thus occur on these scales, $C(k,t)=1$, for 
$kt\ll1$. As soon as defects enter the horizon, they start evolving in a 
complicated non-linear way and their gravitational potential 
looses coherence with a characteristic time scale $t_c$. On the other hand, 
the defects themselves decay with a decay time $t_d$. Once the defects 
have decayed, dark matter and radiation fluctuations evolve according to 
homogeneous perturbation equations not loosing any remaining coherence. 
The question whether decoherence is effective or not, is thus determined 
by the ratio $r=t_c/t_d$. For $r\leq 1$ decoherence is unimportant where 
as for $r\gg 1$ decoherence smears out secondary acoustic peaks leaving 
over just one broad 'hump' \cite{joao}. This process is illustrated in 
Fig.~5.  
\begin{figure}[htb]
\epsfysize=9.5cm
\centerline{\epsffile{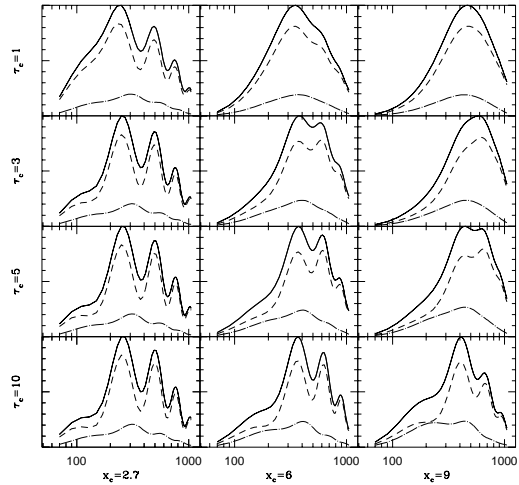}}
\caption{CMB anisotropies from topological defects. The 
effect of decoherence is shown. The variables $\tau_c$ and $x_c$ correspond to
$t_d$ and $t_c$ respectively. The ratio $r$ is thus largest for the
top right frame and smallest for the bottom left one. 
(From Magueijo et al., Ref.~3.)}
\end{figure}

Local cosmic strings decay only via the very weak process of gravitational 
radiation and are thus relatively long lived. A cosmic string loop, after 
entering the horizon, typically  survives for about $10^4$ horizon times.
Global defects, on the other hand, decay very effectively within a 
few horizon times via the radiation of massless Goldstone modes. 
Furthermore, there are hints from numerical simulations, that coherence 
decays exponentially for cosmic strings\cite{joao} but only like a 
power law for global defects\cite{KD}. These findings have led to the 
conjecture that decoherence is effective in scenarios with local cosmic 
strings but not for global defects. However, to fully understand and 
quantify decoherence, more detailed simulations and analytical work are 
needed.

\section{Conclusion}
We have seen that the anisotropies in the CMB provide  interesting 
possibilities to distinguish between models of structure formation from 
inflationary perturbations or from topological defects. The first tests 
for physical theories at very high energies $\sim 10^{16}$GeV might thus 
come from cosmology and not from accelerator experiments! 

One of the reasons for its usefulness certainly lies in the
simplicity of calculations of CMB 
anisotropies. All the effects discussed here can be determined within 
linear perturbation analysis. The complicated non-linear physics involved 
in the formation of celestial bodies plays only a minor role for CMB 
anisotropies, while it might significantly obscure the relation between 
observations and calculations of large scale structure. There, 
the quantities simple to observe are the clustering properties of
light, while linear perturbation analysis 
just determines the clustering  of mass.

We have justified hopes that the next decade, when experimental 
results  determine CMB anisotropy to an accuracy of a few percent, will 
revolutionize cosmology. On the one hand, the dependence of the
details of the acoustic peaks on cosmological parameters\cite{SiHu}
might help us to determine these parameters to an accuracy of a few
percent. On the other hand, the acoustic peaks probably contain
information about the physics at GUT scales
which is not available to us by any other means.

I have not discussed here the distinction of inflationary and defect models 
by statistical means: While generic inflationary models lead to Gaussian 
perturbations, topological defects are inherently non-Gaussian. It may 
however be quite difficult to detect this deviation from Gaussian 
statistics: On very large scales, a significant obstacle is cosmic variance, 
while on intermediate scales, several defects might contribute to a 
given perturbation and thus reduce non-Gaussian signatures (central 
limit theorem). The best prospects are probably on small scales, where 
one might actually 'see' the discontinuity due to one cosmic 
string\cite{KS}. An interesting discussion of the problem of statistics 
is given in the contribution by J. Magueijo in these proceedings.

\section*{Acknowledgments}
It is a pleasure to thank Monique Signoret and Francesco Melchiorri for 
organizing this short but stimulating and active meeting. I also want to 
express my thanks to Mairi Sakellariadou who 
contributed to much of the original work reported here. I gratefully 
acknowledge support by the Fonds National Suisse.

\section*{Appendix}
In this appendix we show in some detail how Eqs.~(\ref{k1CDM}) 
and (\ref{defects}) lead both to a Harrison Zel'dovich spectrum of
microwave background anisotropies, {\em i.e.} $C_\ell\propto
1/\ell(\ell+1)$.
 
Using ${\bf x}_{dec}={\bf x}_0-{\bf n}(t_0-t_{dec})\approx 
{\bf x}_0-{\bf n}t_0$, the Fourier 
transform of Eq.~(\ref{CDM}) yields
\be
{\delta T\over T}({\bf n,k})={1\over 3}\exp(i{\bf n\cdot k}t_0)
		\Psi(t_{dec},{\bf k})  ~. \label{kCDM}
\ee
Using the well-known identity\cite{AS}
$\exp(i\al\mu)=\sum_\ell(2\ell+1)i^\ell j_\ell(\al)P_\ell(\mu)$, where
$j_\ell$ denotes the spherical Bessel function and $P_\ell$ is the
Legendre polynomial of index $\ell$, we
find
\be
{\delta T\over T}(k,\mu)=\sum_\ell i^\ell \De_\ell(k)P_\ell(\mu)~, 
\ee
with $\mu={\bf n\cdot k}/k$, and
\be
\De_\ell(k)={2\ell +1\over 3}\Psi(k,t_{dec})j_\ell(kt_0) ~.
 \label{Deinf}\ee
Inserting this in the two point correlation function,
Eq.~(\ref{2point}), one obtains
\be
  C_\ell = {2\over \pi}\int dk k^2{\langle|\De(k)|^2\rangle\over(2\ell+1)^2} 
~.\label{lCDM}
\ee
To arrive at this result we replace the ensemble average of
Eq.~(\ref{2point}) by an integration over observer positions ${\bf
x}_0$, a kind of 'ergodic hypothesis'. Then we  use unitarity of
the Fourier transform and elementary orthogonality properties of
spherical harmonics. The average $\langle|\De(k)|^2\rangle$ represents
an integral over $\bf k$-directions. For Eq.~(\ref{lCDM}) to hold, it
is important that the Fourier transform is defined by
\be \hat{f}({\bf k}) = {1\over\sqrt{V}}\int_V d^3xf({\bf x})
	\exp(i{\bf k\cdot x})~,
\ee
otherwise, the pre-factor in front of the integral in Eq.~(\ref{lCDM})
changes.

Inserting now $\langle|\Psi|^2(k)\rangle=Ak^{-3}$, the integral in
Eq~(\ref{lCDM}) becomes
\[
C_\ell= {2A\over9\pi}\int {dx\over x}j_\ell(x)^2 ~.
\]
This integral can be performed exactly with the result\cite{GR}
\be
C^{(inflation)}_\ell = {A\Ga(\ell)\over 36\Ga(3/2)\Ga(\ell+2)}
	={A\over9\pi\ell(\ell+1)} 
~. \label{Clinf}\ee
The dimensionless constant $A$ is given by the specific inflationary
model and has to be tuned to $A\sim 10^{-9}$.

For topological defects the situation is somewhat more complicated
since $\Psi$ is time dependent. 
The same reasoning which led to Eq.~(\ref{Deinf}) yields  
\be {1\over 2\ell+1}\De_\ell(k)=
2\Psi(t_{dec},k)j_\ell(kt_0) + 2\int_{t_{dec}}^{t_0}
	\dot{\Psi}(t,k)j_\ell(k(t_0-t))dt~. \label{Dede}
\ee
If we now make use of the approximation for the geometry perturbations
from defects given in Eq.~(\ref{defects}) and simply set $\Psi({\bf k})\sim
\sqrt{\langle|\Psi(k)|^2\rangle}$ , the integration in
Eq.~(\ref{Dede}) has to be
performed only until $t=1/k$, since $\dot{\Psi}$ vanishes for
$kt>1$. We thus may neglect the weak time dependence of $j_\ell$ in
the interval of integration. The
integral in Eq.~(\ref{Dede}) can then be performed. The ordinary SW
contribution is canceled by the lower limit of the integral and only
the much larger contribution from the upper limit of the integral
remains leading to
\be \De_\ell(k)=2(2\ell+1)\Psi(t=1/k,k)j_\ell(kt_0) ~. \ee
Inserting this finding in Eq.~(\ref{lCDM}) with 
$\langle|\Psi(t=1/k,k)|^2\rangle=A\ep^2k^{-3}$, we find as 
in Eq.~(\ref{Clinf})
\be C_\ell^{(\defect)} = {8A\ep^2\over\pi\ell(\ell+1)} ~.\ee
Here $A$ is a dimensionless constant which is model dependent, but
generically of order unity, and $\ep=4\pi G\eta^2$ is given by the
symmetry breaking scale. 

Even if the integration encountered in this derivation can be
performed exactly, we should not forget that the crucial ingredient,
Eq.~(\ref{defects}) is a  simple approximation and the constant
$A$ should be determined by numerical simulations. Also in the simple
inflationary CDM model, massless neutrini and the contribution of
radiation to the expansion of the universe induce a small integrated
SW contribution which has been neglected in this approximation. 

In general, if $\langle|\De_\ell(k)|^2\rangle = Ak^{-n}t_0^{3-n}$ in a
certain range of harmonics $\ell$, one can
compute the $C_\ell$'s in this range exactly with the result
\[ C_\ell= A{\Ga(n-1)\Ga(\ell+{3-n\over 2})\over
2^{n-1}\Ga(n/2)^2\Ga(\ell+{n+1\over 2})} ~.\]

\end{document}